\documentclass{JHEP3}%
\usepackage{amsmath}
\usepackage{mathbbol}
\usepackage{graphicx}
\usepackage{amssymb}
\usepackage{epsfig,multicol,bbm}
\usepackage{amsfonts}%
\providecommand{\U}[1]{\protect\rule{.1in}{.1in}}
\graphicspath{{E:/cygwin/home/Forshaw/MyPapers/SuperLeading/}}
\newcommand{\mytitle}{\title}
\newcommand{\myauthor}{\author}
\mytitle{Super-leading logarithms in non-global observables in QCD?}
\myauthor{J.R. Forshaw, A. Kyrieleis \\School of Physics \& Astronomy, University of Manchester, \\
Oxford Road, Manchester M13 9PL, U.K.\\
\email{forshaw@mail.cern.ch}, \email{kyrieleis@hep.man.ac.uk}}
\myauthor{M.H. Seymour \\ School of Physics \& Astronomy, University of Manchester and
Theoretical Physics Group (PH-TH), CERN, CH-1211 Geneva 23, Switzerland \\ \email{mike.seymour@cern.ch}}
\preprint{MAN/HEP/2006/1 \\ CERN--PH--TH/2006--058}
\abstract{We reconsider the calculation of a non-global QCD observable and find the possible breakdown of
QCD coherence. This breakdown arises as a result of wide angle soft gluon emission developing a sensitivity to
emission at small angles and it leads to the appearance of super-leading logarithms. We use the `gaps between jets' cross-section as a concrete example and illustrate that the new logarithms are intimately connected with the presence of Coulomb gluon contributions. We present some rough estimates of their potential phenomenological significance.}
\ifx\pdfoutput\relax\let\pdfoutput=\undefined\fi
\newcount\msipdfoutput
\ifx\pdfoutput\undefined\else
\ifcase\pdfoutput\else
\ifx\paperwidth\undefined\else
\ifdim\paperheight=0pt\relax\else\pdfpageheight\paperheight\fi
\ifdim\paperwidth=0pt\relax\else\pdfpagewidth\paperwidth\fi
\fi\fi\fi
\begin{document}
\section{Introduction}

The summation of single logarithmic effects in QCD observables arising as a
consequence of `wide angle' soft gluon emission has a long history
\cite{Collins88}--\cite{Kyrieleis:2005dt}, with the discovery of non-global
logarithms providing a recent highlight \cite{DasSa1}\cite{DasSa2}. In this
paper, we wish to report the possible emergence of a new class of
`super-leading' logarithms which could arise in general non-global
observables. We refer to the logarithms as super-leading since they are
formally more important than the `leading logarithmic' summations that have
hitherto been performed. The fact that these new contributions first arise at
quite a high order in the perturbative expansion in processes involving at
least four external coloured particles and are subleading in the number of
colours $N$ may account for their lying undiscovered until now. Their origin
is related to the non-Abelian Coulomb phase terms which are present in the
colour evolution.

At the present stage in our understanding, we are not able to claim strictly
to have proven the existence of super-leading logarithms and the corresponding
breakdown of QCD coherence that such logarithms would imply. As we shall see,
the superleading logarithms emerge under the assumption that successive
emissions can be ordered in transverse momentum and we have not proven
this\footnote{We expect similar logarithms to emerge using other ordering
variables although it is possible that the coefficient of the superleading
logarithm may differ.}. We do however wish to stress that the failure of
$k_{T}$ ordering would itself be of significant interest.

The paper is organized as follows. Throughout we shall focus upon one
particular non-global observable, namely the `gaps between jets'
cross-section, although our conclusions are clearly more general. This is the
cross-section for producing a pair of high transverse momentum jets ($Q$) with
a restriction on the transverse momentum of any additional jets radiated in
between the two leading jets, i.e.\ $k_{T}<Q_{0}$ for emissions in the gap
region. This process has been much studied in the literature
\cite{Oderda:1998en}--\cite{Appleby:2003sj} and has been measured
experimentally \cite{Derrick:1995pb}--\cite{Abe:1998ip}. In the following
section we explain how to sum the logarithms which arise as a result of soft
gluon corrections to the hard scattering. We explain how the non-global nature
of the observable affects the summation and in particular how it necessitates
the summation over real and virtual soft gluon emissions outside of the region
between the jets. We organize our calculation in terms of the number of gluons
which lie outside of the gap region and compute the contribution to the
cross-section which arises from one emission (real or virtual) outside of the
gap. In Section \ref{sec:SLL}, we uncover the super-leading logarithmic
structure. We show that it is intimately connected with the imaginary ($i\pi$)
terms which are present in the soft gluon evolution due to the exchange of
Coulomb gluons, and that it is subleading in $N$. We show also that the
super-leading logarithms can be seen to arise as a result of the breakdown of
the `plus prescription' in the evolution of radiation which is collinear with
either of the incoming parton legs above the scale $Q_{0}$ because, although
the real and virtual parts (at $z<1$ and $z=1$ respectively) are equal and
opposite, their subsequent evolution down to the scale $Q_{0}$ is not. In
Section \ref{sec:results} we present some numerical results.

\section{Some history}

In the original calculations of the gaps between jets cross-section
\cite{Oderda:1998en}\cite{Oderda:1999kr}, all those terms $\sim\alpha_{s}%
^{n}\ln^{n}(Q/Q_{0})$ that can be obtained by dressing the primary
$2\rightarrow2$ scattering in all possible ways with soft virtual gluons were
summed. The restriction to soft gluons implies the use of the eikonal
approximation and greatly simplifies the calculations. For the validity of the
eikonal approximation, it is assumed that all collinear radiation can be
summed inclusively and hence that any collinear logarithms in $Q/Q_{0}$ can be
absorbed into the incoming parton density functions. We shall later question
this assumption but for now we assume its validity.

We shall focus our attention upon quark-quark scattering: the colour structure
is simpler and all of the key ideas are present. In this case, the re-summed
scattering amplitude can be written\footnote{We neglect the running of the
strong coupling throughout this paper although it is straightforward to
re-instate it.}%
\begin{equation}
\mathbf{M}(Q_{0})=\exp\left(  -\frac{2\alpha_{s}}{\pi}%
{\displaystyle\int\limits_{Q_{0}}^{Q}}
\frac{dk_{T}}{k_{T}}\mathbf{~\Gamma}\right)  \mathbf{M}(Q), \label{eq:OS}%
\end{equation}
where \cite{Sotir93}\cite{Kidonakis:1998nf}
\begin{equation}
\mathbf{\Gamma}=\left(
\begin{array}
[c]{cc}%
\frac{N^{2}-1}{4N}\rho(Y,\Delta y) & \frac{N^{2}-1}{4N^{2}}i\pi\\
i\pi & -\frac{1}{N}i\pi+\frac{N}{2}Y+\frac{N^{2}-1}{4N}\rho(Y,\Delta y)
\end{array}
\right)  \label{eq:gamma0}%
\end{equation}
is the matrix which tells us how to attach a soft gluon to a primary
four-quark hard scattering. $\mathbf{\Gamma}$ is defined in the $t$-channel
(singlet-octet) basis where
\begin{equation}
\sigma=\mathbf{M}^{\dag}\mathbf{S}_{V}\mathbf{M}%
\end{equation}
is the scattering cross-section and
\begin{equation}
\mathbf{M=}\left(
\begin{array}
[c]{c}%
M^{(1)}\\
M^{(8)}%
\end{array}
\right)  ~\text{and }\mathbf{S}_{V}\mathbf{=}\left(
\begin{array}
[c]{cc}%
N^{2} & 0\\
0 & \frac{N^{2}-1}{4}%
\end{array}
\right)  .
\end{equation}
In Eq.(\ref{eq:gamma0}), $Y$ is the size of the rapidity region over which
emission with $k_{T}>Q_{0}$ is vetoed and $\Delta y$ is the distance between
the two jet centres (in the most commonly used event definition, $\Delta
y=Y+2R$ \ where $R$ is the radius of the jet cone) and%
\begin{equation}
\rho(Y,\Delta y)=\log\frac{\sinh(\Delta y/2+Y/2)}{\sinh(\Delta y/2-Y/2)}-Y.
\end{equation}
In this basis we have%
\begin{equation}
\mathbf{M}(Q)\equiv\mathbf{M}_{0}=\sqrt{\frac{4}{N^{2}-1}}\left(
\begin{array}
[c]{c}%
0\\
\sqrt{\sigma_{\text{born}}}%
\end{array}
\right)  .
\end{equation}
Equation (\ref{eq:gamma0}) quantifies the effect of adding a soft and virtual
gluon in all possible ways to a four-quark matrix element. Within the eikonal
approximation one obtains contributions from two distinct regions of the loop
integral: the first, sometimes denoted the `eikonal gluon' contribution
\cite{Dokshitzer:2005ig}, comes from the pole at $k^{2}=0$. The residue of
this pole is identical, but with opposite sign, to the phase space integral
for the emission of a real soft gluon. In particular, it makes sense to
ascribe definite values of rapidity, azimuth and transverse momentum to the
eikonal virtual gluon. Of these, Eq.(\ref{eq:gamma0}) includes only those
whose rapidity lies within the gap region (i.e.\ within the region between the
two hard jets), the contributions from outside the gap region cancelling with
corresponding real emission contributions. The second region of the loop
integral, sometimes denoted the `Coulomb gluon' contribution
\cite{Dokshitzer:2005ig,Ralston:1982pa,Pire:1982iv}, comes from a pole at
which one of the emitting partons is on-shell, which pinches the contour
integrals at the point at which the gluon's positive and negative light-cone
momenta are zero, corresponding to a space-like gluon with only transverse
momentum. Its residue is purely imaginary and only non-zero if both the
emitting partons are in the final state or both in the initial state, giving
the $i\pi$ terms in the evolution matrix. Strictly speaking, for the Coulomb
gluons, the region of $k_{T}$ below $Q_{0}$ must be included, however this
region only contributes a pure phase which cancels in observables, as shown
explicitly in \cite{Dokshitzer:2005ig}.

At first sight, one may suppose that the evolution just described correctly
captures all of the leading logarithms. This would indeed be so if it were the
case that the contributions arising from real gluon emission always cancel
with a corresponding virtual emission. In this case, the only region in
phase-space where the real-virtual cancellation would not occur would be the
region where real emissions are forbidden, i.e.\ within the gap region and
with transverse momentum above $Q_{0}$. This may seem a straightforward
consequence of the Bloch-Nordsieck Theorem however it is not.

\begin{figure}[h]
\begin{center}
\includegraphics[width=4in]{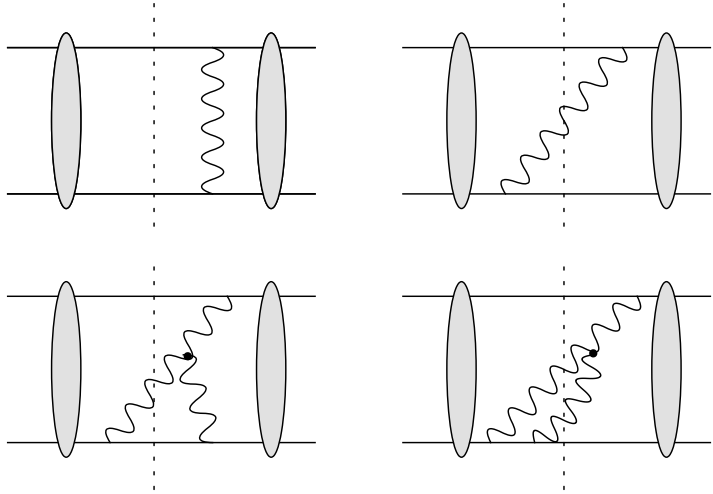}\\[0pt]
\end{center}
\caption{Illustrating the cancellation (and miscancellation) of soft gluon
corrections.}%
\label{fig:miscancel}%
\end{figure}

Although it is true that the real and virtual contributions cancel exactly at
the cross-section level in the case where we dress a hard scattering amplitude
with a single soft gluon, it is not true that the cancellation survives
subsequent dressing with additional soft gluons. This is illustrated in
Fig.\ref{fig:miscancel}. The upper two panes contain typical diagrams where a
soft gluon dresses a $2\rightarrow2$ hard scattering. At the cross-section
level such single soft gluon corrections exactly cancel each other since all
cuts through a particular uncut diagram sum to zero. Now consider the lower
two panes. To capture the leading logarithms we assume that it is appropriate
to order strongly the transverse momenta of successive gluon emissions as one
moves away from the hard scatter. If we first consider a real gluon emission
above $Q_{0}$ then it must lie outside of the gap region. We should then
consider virtual corrections to this five parton amplitude. Bloch-Nordsieck
guarantees only that it is true that those virtual corrections which lie
outside the gap region in rapidity, or have transverse momentum below $Q_{0}$,
will be exactly cancelled by the corresponding real emission graphs. Virtual
corrections to the five-parton amplitude which lie above $Q_{0}$ and are
within the gap region have nothing to cancel against, for their corresponding
real emissions are forbidden by the definition of the observable. These
virtual corrections embody the fact that any emission outside of the gap
region is forbidden from radiating back into the gap with $k_{T}>Q_{0}$. Thus
we see that the non-global nature of the observable has prevented the soft
gluon cancellation which is necessary in order that Eq.(\ref{eq:OS}) should be
the whole story.

It is therefore necessary to include the emission of any number of soft gluons
outside the gap region (real and virtual) dressed with any number of virtual
gluons within the gap region; all gluons having transverse momentum above
$Q_{0}$. Clearly it is a formidable challenge to sum all of the leading
logarithms, mainly because of the complicated colour structure of an amplitude
with a large number of final state gluons. Progress has been made, working
within the large $N$ approximation \cite{Appleby:2002ke,Appleby:2003sj}. In
fact a great deal of interest has been generated \cite{Marchesini:2003nh}%
--\cite{Marchesini:2004ne} by the fact that, in this large $N$ limit, the
evolution equation for the out-of-gap gluons \cite{Banfi:2002hw} maps onto the
Kovchegov equation for non-linear corrections to the BFKL equation
\cite{Kovchegov:1999ua}--\cite{Weigert:2005us}. Here, however, we prefer to
keep the exact colour structure but instead we only compute the cross-section
for one gluon outside of the gap region. This can be viewed as the first term
in an expansion in the number of out-of-gap gluons.

\subsection{One emission outside of the gap}

Thus motivated, we now compute the cross-section for emitting one soft gluon
outside of the gap region dressed with any number of virtual gluons. There are
two new ingredients compared to the four-parton case:

\begin{enumerate}
\item We need to consider the emission of a real gluon off any one of the four
external quarks. The corresponding five parton amplitude needs four colour
basis states and hence the action of emitting a real gluon from the four-quark
amplitude will be described using a 4x2 matrix, $\varepsilon^{\mu}%
\mathbf{D}_{\mu}$, where $\varepsilon^{\mu\text{ }}$denotes the gluon
polarisation vector.

\item We need also to determine the 4x4 matrix\textbf{\ }$\mathbf{\Lambda}%
$\textbf{\ }which acts on the five particle amplitude in order to account for
the dressing with a virtual soft gluon.
\end{enumerate}

The real emission contribution is obtained from the four-quark amplitude
$\mathbf{M}$ via\footnote{For notational convenience, we suppress the
dependence on the rapidity and azimuth of the emitted gluon.}%
\begin{equation}
\mathbf{M}_{R}^{\mu}(k_{T})=\mathbf{D}^{\mu}\mathbf{M(}k_{T})
\end{equation}
with%
\begin{equation}
\mathbf{D}^{\mu}=\left(
\begin{array}
[c]{cc}%
\frac{1}{2}(-h_{1}^{\mu}-h_{2}^{\mu}+h_{3}^{\mu}+h_{4}^{\mu}) & \frac{1}%
{4N}(-h_{1}^{\mu}-h_{2}^{\mu}+h_{3}^{\mu}+h_{4}^{\mu})\\
0 & \frac{1}{2}(-h_{1}^{\mu}-h_{2}^{\mu}+h_{3}^{\mu}+h_{4}^{\mu})\\
\frac{1}{2}(-h_{1}^{\mu}+h_{2}^{\mu}+h_{3}^{\mu}-h_{4}^{\mu}) & \frac{1}%
{4N}(h_{1}^{\mu}-h_{2}^{\mu}-h_{3}^{\mu}+h_{4}^{\mu})\\
0 & \frac{1}{2}(-h_{1}^{\mu}+h_{2}^{\mu}-h_{3}^{\mu}+h_{4}^{\mu})
\end{array}
\right)  \label{eq:tchannelD}%
\end{equation}
and the eikonal factors are%
\begin{equation}
h_{i}^{\mu}=\frac{1}{2}k_{T}\frac{p_{i}^{\mu}}{p_{i}\cdot k},
\end{equation}
where $k$ is the gluon's four-momentum and $p_{i}$ are the external quark
momenta. In particular, we choose%
\begin{align}
p_{1}  &  =\frac{\sqrt{s}}{2}\left(  1;0,0,1\right)  ,\nonumber\\
p_{2}  &  =\frac{\sqrt{s}}{2}\left(  1;0,0,-1\right)  ,\nonumber\\
p_{3}  &  =Q\left(  \cosh(\Delta y/2);0,1,\sinh(\Delta y/2)\right)
,\nonumber\\
p_{4}  &  =Q\left(  \cosh(\Delta y/2);0,-1,-\sinh(\Delta y/2)\right)
,\nonumber\\
k  &  =k_{T}\left(  \cosh y;\sin\phi,\cos\phi,\sinh y\right)  .
\label{eq:fourmomenta}%
\end{align}
Eq.(\ref{eq:tchannelD}) is defined in the `$t$-channel' basis, i.e.\ the four
basis vectors for the process $q_{i}q_{j}\rightarrow q_{k}q_{l}g_{a}$ are%
\begin{align}
\mathbf{C}_{1}  &  =T_{ki}^{a}\delta_{lj}+T_{lj}^{a}\delta_{ki},\\
\mathbf{C}_{2}  &  =T_{ki}^{b}T_{lj}^{c}~d^{abc},\\
\mathbf{C}_{3}  &  =T_{ki}^{a}\delta_{lj}-T_{lj}^{a}\delta_{ki},\\
\mathbf{C}_{4}  &  =T_{ki}^{b}T_{lj}^{c}~if^{abc}.
\end{align}
The cross-section for one real gluon emission off the four-quark amplitude
$\mathbf{M}$ is then given by\footnote{The minus sign arises after the sum
over gluon polarisations using $\sum\varepsilon_{\mu}^{\ast}\varepsilon_{\nu
}=-g_{\mu\nu}$.}%
\begin{equation}
\sigma_{R}=-\frac{2\alpha_{s}}{\pi}\int\frac{dk_{T}}{k_{T}}\int\frac{dy~d\phi
}{2\pi}(\mathbf{M}^{\dag}\mathbf{D}_{\mu}^{\dagger}\mathbf{S}_{R}%
\mathbf{D}^{\mu}\mathbf{M)}%
\end{equation}
where%
\begin{equation}
\mathbf{S}_{R}\mathbf{=}\left(
\begin{array}
[c]{cccc}%
N(N^{2}-1) & 0 & 0 & 0\\
0 & \frac{1}{4N}(N^{2}-1)(N^{2}-4) & 0 & 0\\
0 & 0 & N(N^{2}-1) & 0\\
0 & 0 & 0 & \frac{1}{4}N(N^{2}-1)
\end{array}
\right)  .
\end{equation}
One can readily check that the single soft gluon cancellation is assured
since\footnote{The cancellation occurs already at the level of the integrand.}%
\begin{equation}
\int_{\text{gap}}\frac{dy~d\phi}{2\pi}\mathbf{D}_{\mu}^{\dag}\mathbf{S}%
_{R}\mathbf{D}^{\mu}+\mathbf{\Gamma}^{\dag}\mathbf{S}_{V}+\mathbf{S}%
_{V}\mathbf{\Gamma}=\mathbf{0}.
\end{equation}
The subsequent evolution of this five parton amplitude is determined by
$\mathbf{\Lambda}$:
\begin{equation}
\mathbf{M}_{R}(Q_{0})=\exp\left(  -\frac{2\alpha_{s}}{\pi}%
{\displaystyle\int\limits_{Q_{0}}^{k_{T}}}
\frac{dk_{T}^{\prime}}{k_{T}^{\prime}}~\mathbf{\Lambda}\right)  \mathbf{M}%
_{R}(k_{T}),
\end{equation}
where the evolution matrix was computed in \cite{Kyrieleis:2005dt} to be%
\begin{align}
\hspace*{-3cm}\mathbf{\Lambda}  &  =\left(
\begin{array}
[c]{cccc}%
\frac{N}{4}{\normalsize (Y-i\pi)+}\frac{1}{2N}{\normalsize i\pi} & \left(
\frac{1}{4}-\frac{1}{N^{2}}\right)  {\normalsize i\pi} & {\normalsize -}%
\frac{N}{4}{\normalsize s}_{y}{\normalsize Y} & {\normalsize 0}\\
{\normalsize i\pi} & \frac{N}{4}\left(  2Y-i\pi\right)  {\normalsize -}%
\frac{3}{2N}{\normalsize i\pi} & {\normalsize 0} & {\normalsize 0}\\
{\normalsize -}\frac{N}{4}{\normalsize s}_{y}{\normalsize Y} & {\normalsize 0}
& \frac{N}{4}{\normalsize (Y-i\pi)-}\frac{1}{2N}{\normalsize i\pi} &
{\normalsize -}\frac{1}{4}{\normalsize i\pi}\\
{\normalsize 0} & {\normalsize 0} & {\normalsize -i\pi} & \frac{N}{4}\left(
2Y-i\pi\right)  {\normalsize -}\frac{1}{2N}{\normalsize i\pi}%
\end{array}
\right) \nonumber\\
&  +\left(
\begin{array}
[c]{cccc}%
N & 0 & 0 & 0\\
0 & N & 0 & 0\\
0 & 0 & N & 0\\
0 & 0 & 0 & N
\end{array}
\right)  \frac{1}{4}\rho(Y,2\left\vert y\right\vert )\nonumber\\
&  +\left(
\begin{array}
[c]{cccc}%
C_{F} & 0 & 0 & 0\\
0 & C_{F} & 0 & 0\\
0 & 0 & C_{F} & 0\\
0 & 0 & 0 & C_{F}%
\end{array}
\right)  \frac{1}{2}\rho(Y,\Delta y)\nonumber\\
&  +\left(
\begin{array}
[c]{cccc}%
-\frac{N}{4} & 0 & -\frac{N}{4}s_{y} & \frac{1}{4}s_{y}\\
0 & -\frac{N}{4} & 0 & \frac{N}{4}s_{y}\\
-\frac{N}{4}s_{y} & 0 & -\frac{N}{4} & -\frac{1}{4}\\
s_{y} & \left(  \frac{N}{4}-\frac{1}{N}\right)  s_{y} & -1 & -\frac{N}{4}%
\end{array}
\right)  \frac{1}{2}\lambda
\end{align}
with
\begin{align}
\lambda &  =\frac{1}{2}\log\frac{\cosh(\Delta y/2+\left\vert y\right\vert
+Y)-s_{y}\cos(\phi)}{\cosh\left(  \Delta y/2+\left\vert y\right\vert
-Y\right)  -s_{y}\cos(\phi)}-Y,\\
s_{y}  &  =\text{sgn}(y).
\end{align}

We now have the machinery to state the all-orders cross-section for one gluon
outside of the gap. For the real emission we have%
\begin{align}
\sigma_{R}  &  =-\frac{2\alpha_{s}}{\pi}\int_{Q_{0}}^{Q}\frac{dk_{T}}{k_{T}%
}~\int\limits_{\text{out}}\frac{dy~d\phi}{2\pi}~\nonumber\\
&  \mathbf{M}_{0}^{\dag}\exp\left(  {\small -}\frac{2\alpha_{s}}{\pi}%
\int\limits_{k_{T}}^{Q}\frac{dk_{T}^{\prime}}{k_{T}^{\prime}}\mathbf{\Gamma
}^{\dag}\right)  \mathbf{D}_{\mu}^{\dag}\exp\left(  {\small -}\frac
{2\alpha_{s}}{\pi}\int\limits_{Q_{0}}^{k_{T}}\frac{dk_{T}^{\prime}}%
{k_{T}^{\prime}}\mathbf{\Lambda}^{\dag}\right)  \mathbf{S}_{R}\nonumber\\
&  \exp\left(  {\small -}\frac{2\alpha_{s}}{\pi}\int\limits_{Q_{0}}^{k_{T}%
}\frac{dk_{T}^{\prime}}{k_{T}^{\prime}}\mathbf{\Lambda}\right)  \mathbf{D}%
^{\mu}\exp\left(  {\small -}\frac{2\alpha_{s}}{\pi}\int\limits_{k_{T}}%
^{Q}\frac{dk_{T}^{\prime}}{k_{T}^{\prime}}\mathbf{\Gamma}\right)
\mathbf{M}_{0} \label{eq:real}%
\end{align}
and for a virtual emission%
\begin{align}
\sigma_{V}  &  =-\frac{2\alpha_{s}}{\pi}\int_{Q_{0}}^{Q}\frac{dk_{T}}{k_{T}%
}\int\limits_{\text{out}}\frac{dy~d\phi}{2\pi}\nonumber\\
&  \left[  \mathbf{M}_{0}^{\dag}\exp\left(  {\small -}\frac{2\alpha_{s}}{\pi
}\int\limits_{Q_{0}}^{Q}\frac{dk_{T}^{\prime}}{k_{T}^{\prime}}\mathbf{\Gamma
}^{\dag}\right)  \mathbf{S}_{V}\right. \nonumber\\
&  \left.  \exp\left(  {\small -}\frac{2\alpha_{s}}{\pi}\int\limits_{Q_{0}%
}^{k_{T}}\frac{dk_{T}^{\prime}}{k_{T}^{\prime}}\mathbf{\Gamma}\right)
~\boldsymbol{\gamma}\mathbf{~}\exp\left(  {\small -}\frac{2\alpha_{s}}{\pi
}\int\limits_{k_{T}}^{Q}\frac{dk_{T}^{\prime}}{k_{T}^{\prime}}\mathbf{\Gamma
}\right)  \mathbf{M}_{0}~{\small +~}\text{{\small c.c.}}\right]
\label{eq:virtual}%
\end{align}
where the matrix $\boldsymbol{\gamma}$\textbf{\ }adds the virtual soft gluon
which is to lie outside of the gap. It differs from $\mathbf{\Gamma}$ in the
fact that the rapidity integral is left undone and in that it is purely real
since the imaginary Coulomb terms have already been entirely accounted for by
the evolution matrix $\mathbf{\Gamma}$. We have that%
\begin{equation}
\boldsymbol{\gamma}=\frac{1}{2}\left(
\begin{tabular}
[c]{cc}%
$\frac{N^{2}-1}{2N}\left(  \omega_{13}+\omega_{24}\right)  $ & $\frac{N^{2}%
-1}{4N^{2}}\left(  -\omega_{12}-\omega_{34}+\omega_{14}+\omega_{23}\right)
\medskip$\\
$-\omega_{12}-\omega_{34}+\omega_{14}+\omega_{23}\;$ & $\;%
\begin{array}
[t]{c}%
\frac{N}{2}(\omega_{14}+\omega_{23})-\frac{1}{2N}\left(  \omega_{13}%
+\omega_{24}\right) \\
\qquad+\frac{1}{N}\left(  \omega_{12}+\omega_{34}-\omega_{14}-\omega
_{23}\right)
\end{array}
$%
\end{tabular}
\right)  \label{eq:gamma}%
\end{equation}
\newline where%
\begin{equation}
\omega_{ij}\equiv2h_{i}\cdot h_{j}=\frac{1}{2}k_{T}^{2}\frac{p_{i}\cdot p_{j}%
}{(p_{i}\cdot k)(p_{j}\cdot k)}.
\end{equation}

\section{Super-leading logarithms\label{sec:SLL}}

In the next section we shall present some numerical results obtained by
evaluating the sum of equations (\ref{eq:real}) and (\ref{eq:virtual}) but
first we shall take a closer look at the singularity structure of each. We
expect both to contain divergences in the formal limit that the out-of-gap
gluon becomes collinear with any of the external quarks and we might suppose
that these divergences always cancel. Such cancellations are to be expected as
a result of QCD coherence which informs us that large angle soft gluon
emission should not be able to resolve emissions at small angles. Let us first
explore emissions that are collinear with an outgoing quark.

\subsection{Final state collinear emission}

We state the result first: emissions collinear to an outgoing quark do cancel
between the real and virtual corrections. To see this it is better to shift to
a colour basis in which the evolution matrix $\mathbf{\Lambda}$ is block
diagonal. The relevant results are summarized in Appendix \ref{app2}. Let's
consider the particular case in which the emission is collinear with $p_{3}$
(i.e.\ $y>0$). In this case, Eq.(\ref{eq:gamma}) simplifies to%
\begin{equation}
\boldsymbol{\gamma}\rightarrow\frac{N^{2}-1}{4N}~\omega_{3}\left(
\begin{array}
[c]{cc}%
1 & 0\\
0 & 1
\end{array}
\right)  ,
\end{equation}
where $\omega_{3}=\omega_{13}=\omega_{23}=\omega_{34}$ are the collinear
divergent eikonal factors. Similarly, $\mathbf{D}^{\mu}$ can be much
simplified by keeping only those terms that will induce the collinear
divergence,~i.e.
\begin{equation}
\mathbf{D}^{\mu}=\sqrt{\frac{N^{2}-1}{2N}}(h_{3}^{\mu}-h^{\mu})\left(
\begin{array}
[c]{cc}%
0 & 0\\
0 & 0\\
1 & 0\\
0 & 1
\end{array}
\right)  ,
\end{equation}
where we have taken $h_{1}=h_{2}=h_{4}=h$. Using the fact that in the
collinear limit we should take $\phi=0$, $y=\Delta y/2$ and hence
$\lambda=\rho(Y,2\left\vert y\right\vert )=\rho(Y,\Delta y)$, the evolution is
described by%
\begin{equation}
\mathbf{\Lambda}=\left(
\begin{array}
[c]{cc}%
\begin{array}
[c]{cc}%
\lambda_{1}+\frac{N+1}{4}\rho(Y,\Delta y) & 0\\
0 & \lambda_{2}+\frac{N-1}{4}\rho(Y,\Delta y)
\end{array}
& \mathbf{0}\\
\mathbf{0} & \mathbf{\Gamma}%
\end{array}
\right)  ,
\end{equation}
where the upper left block is unimportant for the evolution because of the
structure of $\mathbf{D}^{\mu}$ (see Eq.(\ref{eq:eigenvals}) for the
definition of $\lambda_{i}$). The final ingredient is the matrix
$\mathbf{S}_{R\text{ }}$ which has the property that its bottom right-hand
entries coincide with the matrix $\mathbf{S}_{V}$, i.e.%
\begin{equation}
\mathbf{S}_{R}=\left(
\begin{array}
[c]{cc}%
\begin{array}
[c]{cc}%
\frac{N^{2}}{2}\frac{N+1}{N+2} & 0\\
0 & \frac{N^{2}}{2}\frac{N-1}{N-2}%
\end{array}
& \mathbf{0}\\
\mathbf{0} & \mathbf{S}_{V}%
\end{array}
\right)  .
\end{equation}
Again the upper left block is not important for the argument here. Hence in
this collinear limit, the evolution of the five-parton amplitude collapses
into the evolution of the four-parton amplitude and we are guaranteed a
complete cancellation between the real and virtual emissions, i.e.\ since
$(h_{3}-h)^{2}=-\omega_{3}$ it follows that
\begin{equation}
\mathbf{D}^{\mu\dag}(\mathbf{\Lambda}^{\dag}\mathbf{)}^{n-m}\mathbf{S}%
_{R}\mathbf{\Lambda}^{m}\mathbf{D}_{\mu}+(\mathbf{\Gamma}^{\dag}%
\mathbf{)}^{n-m}\mathbf{S}_{V}\mathbf{\Gamma}^{m}\boldsymbol{\gamma
}+\boldsymbol{\gamma}^{\dag}(\mathbf{\Gamma}^{\dag}\mathbf{)}^{n-m}%
\mathbf{S}_{V}\mathbf{\Gamma}^{m}=\mathbf{0}.
\end{equation}

\subsection{Initial state collinear emission}

Now we turn our attention to the case where the out-of-gap gluon is collinear
with an incoming quark. It is perhaps worth recalling that by `collinear' we
mean that the rapidity is tending to infinity and $k_{T}>Q_{0}$. Arbitrarily,
we choose the emission to be collinear with $p_{1}$ (i.e.\ $y>0$). Now%
\begin{equation}
\boldsymbol{\gamma}\rightarrow\frac{N^{2}-1}{4N}~\omega_{1}\left(
\begin{array}
[c]{cc}%
1 & 0\\
0 & 1
\end{array}
\right)  ,
\end{equation}
where $\omega_{1}=\omega_{12}=\omega_{13}=\omega_{14}$ are the collinear
divergent eikonal factors in this case. The real emission matrix is not so
simple this time:%
\begin{equation}
\mathbf{D}^{\mu}=\sqrt{\frac{N^{2}-1}{2N}}(h^{\mu}-h_{1}^{\mu})\left(
\begin{array}
[c]{cc}%
0 & \frac{1}{2}\frac{N+2}{N+1}\\
0 & \frac{1}{2}\frac{N-2}{N-1}\\
1 & 0\\
0 & -\frac{1}{N^{2}-1}%
\end{array}
\right)  .
\end{equation}
The evolution matrix $\mathbf{\Lambda\ }$is slightly different too since
$\phi=0$, $y\rightarrow\infty$ in this limit and hence $\lambda=\rho
(Y,2\left\vert y\right\vert )=0$, i.e.%

\begin{equation}
\mathbf{\Lambda}=\left(
\begin{array}
[c]{cc}%
\begin{array}
[c]{cc}%
\lambda_{1} & 0\\
0 & \lambda_{2}%
\end{array}
& \mathbf{0}\\
\mathbf{0} & \mathbf{\Gamma}%
\end{array}
\right)  ,
\end{equation}
where%
\begin{align}
\lambda_{1}  &  =\frac{NY}{2}+\frac{N-1}{2N}i\pi+\frac{N^{2}-1}{4N}%
\rho(Y,\Delta y),\nonumber\\
\lambda_{2}  &  =\frac{NY}{2}-\frac{N+1}{2N}i\pi+\frac{N^{2}-1}{4N}%
\rho(Y,\Delta y). \label{eq:eigenvals}%
\end{align}
Now because of the form of $\mathbf{D}^{\mu}$ the upper left blocks of both
$\mathbf{\Lambda}$ and $\mathbf{S}_{R}$ play a role. Clearly any cancellation
between the real and virtual parts is going to occur only for particular forms
of these blocks. Remarkably, the miscancellation lies wholly in the hands of
the $i\pi$ terms in the evolution matrices, for if we artificially switch
these terms off one finds that%
\begin{equation}
\mathbf{\Lambda}\underset{i\pi\rightarrow0}{=}\frac{NY}{2}\left(
\begin{array}
[c]{cccc}%
1 & 0 & 0 & 0\\
0 & 1 & 0 & 0\\
0 & 0 & 0 & 0\\
0 & 0 & 0 & 1
\end{array}
\right)  +\frac{N^{2}-1}{4N}\rho(Y,\Delta y)\mathbf{1}%
\end{equation}
and%
\begin{equation}
\mathbf{\Gamma}\underset{i\pi\rightarrow0}{=}\frac{NY}{2}\left(
\begin{array}
[c]{cc}%
0 & 0\\
0 & 1
\end{array}
\right)  +\frac{N^{2}-1}{4N}\rho(Y,\Delta y)\mathbf{1.}%
\end{equation}
This diagonal and real structure is sufficient for the cancellation to
operate, i.e.%
\begin{equation}
\mathbf{D}^{\mu\dag}(\mathbf{\Lambda}^{\dag}\mathbf{)}^{n-m}\mathbf{S}%
_{R}\mathbf{\Lambda}^{m}\mathbf{D}_{\mu}+(\mathbf{\Gamma}^{\dag}%
\mathbf{)}^{n-m}\mathbf{S}_{V}\mathbf{\Gamma}^{m}\boldsymbol{\gamma
}+\boldsymbol{\gamma}^{\dag}(\mathbf{\Gamma}^{\dag}\mathbf{)}^{n-m}%
\mathbf{S}_{V}\mathbf{\Gamma}^{m}\underset{i\pi\rightarrow0}{=}\mathbf{0}.
\label{eq:ipito0}%
\end{equation}

The $i\pi$ terms in the evolution arising from Coulomb gluons generally
destroy the cancellation between real and virtual emissions in the case that
the out-of-gap gluon is collinear with one of the incoming partons. In more
familiar terms, we appear to have discovered that the `plus prescription'
employed in the splitting functions for collinear evolution fails for
emissions with transverse momentum above $Q_{0}$. It is particularly
interesting that the miscancellation occurs only once one includes the
imaginary parts in the evolution matrices.

As it stands we have a divergence arising from the integral over the rapidity
of the out-of-gap gluon:%
\begin{equation}
~\int\limits_{\text{out}}\frac{dy~d\phi}{2\pi}~\omega_{1}\sim y_{\text{max}%
}-\frac{Y}{2}. \label{eq:rap}%
\end{equation}
In the soft approximation the integral is divergent which is the signal that
we need to go beyond the soft approximation when considering these emissions.
Strictly speaking we ought to work in the collinear (but not soft)
approximation which means that the integral over rapidity ought to be replaced
by%
\begin{equation}
\int d^{2}k_{T}\int\limits_{\text{out}}dy~~\left.  \frac{d\sigma}{dyd^{2}%
k_{T}}\right\vert _{\text{soft}}\rightarrow\int d^{2}k_{T}\left[
\int\limits^{y_{\text{max}}}dy~\left.  \frac{d\sigma}{dyd^{2}k_{T}}\right\vert
_{\text{soft}}+\int\limits_{y_{\text{max}}}^{\infty}dy~\left.  \frac{d\sigma
}{dyd^{2}k_{T}}\right\vert _{\text{collinear}}\right]  . \label{eq:softcol}%
\end{equation}
In this equation $y_{\text{max}}$ is a matching point between the regions in
which the soft and collinear approximations are used. If $y_{\text{max}}$ is
in the region in which both approximations are valid the dependence on it
should cancel in the sum of the two terms. Now we know that%
\begin{equation}
\int\limits_{y_{\text{max}}}^{\infty}dy~\left.  \frac{d\sigma}{dyd^{2}k_{T}%
}\right\vert _{\text{collinear}}=\int\limits_{y_{\text{max}}}^{\infty
}dy~\left(  \left.  \frac{d\sigma_{\text{R}}}{dyd^{2}k_{T}}\right\vert
_{\text{collinear}}+\left.  \frac{d\sigma_{\text{V}}}{dyd^{2}k_{T}}\right\vert
_{\text{collinear}}\right)
\end{equation}
where the contribution due to real gluon emission is%
\begin{align}
\int\limits_{y_{\text{max}}}^{\infty}dy~\left.  \frac{d\sigma_{\text{R}}%
}{dyd^{2}k_{T}}\right\vert _{\text{collinear}}~  &  =\int\limits_{0}%
^{1-\delta}dz\frac{1}{2}\left(  \frac{1+z^{2}}{1-z}\right)  \frac
{q(x/z,\mu^{2})}{q(x,\mu^{2})}A_{\text{R}}\nonumber\\
&  =\int\limits_{0}^{1-\delta}dz\frac{1}{2}\left(  \frac{1+z^{2}}{1-z}\right)
\left(  \frac{q(x/z,\mu^{2})}{q(x,\mu^{2})}-1\right)  A_{\text{R}}%
+\int\limits_{0}^{1-\delta}dz\frac{1}{2}\frac{1+z^{2}}{1-z}A_{\text{R}}
\label{eq:col}%
\end{align}
and the contribution due to virtual gluon emission is%
\begin{equation}
\int\limits_{y_{\text{max}}}^{\infty}dy~\left.  \frac{d\sigma_{\text{V}}%
}{dyd^{2}k_{T}}\right\vert _{\text{collinear}}~=\int\limits_{0}^{1-\delta
}dz\frac{1}{2}\left(  \frac{1+z^{2}}{1-z}\right)  A_{\text{V}}\text{.}%
\end{equation}
In Eq.(\ref{eq:col}), $q(x,\mu^{2})$ is the parton distribution function for a
quark in a hadron at scale $\mu^{2}$ and momentum fraction $x$. The factors
$A_{\text{R}}$ and $A_{\text{V}}$ contain the $z$ independent factors which
describe the soft gluon evolution and the upper limit on the $z$ integral is
fixed since we require $y>y_{\text{max}}$\footnote{The approximation arises
since we assume for simplicity that $\Delta y$ is large and $\delta$ is small.
This approximation does not affect the leading behaviour and can easily be
made exact if necessary.}:
\begin{equation}
\delta\approx\frac{k_{T}}{Q}\exp\left(  y_{\text{max}}-\frac{\Delta y}%
{2}\right)  .
\end{equation}
We have already established that $A_{\text{R}}+A_{\text{V}}\neq0$ due to
Coulomb gluon contributions to the evolution. If it were the case that
$A_{\text{R}}+A_{\text{V}}=0$ then the virtual emission contribution would
cancel identically with the corresponding term in the real emission
contribution leaving behind a term regularised by the `plus prescription'
(since we can safely take $\delta\rightarrow0$ in the first term of
Eq.(\ref{eq:col})). This term could then be absorbed into the evolution of the
incoming quark parton distribution function by choosing the factorisation
scale to equal the jet scale $Q$.

The miscancellation therefore induces an additional contribution of the form%
\begin{align}
\int\limits_{0}^{1-\delta}dz\frac{1}{2}\left(  \frac{1+z^{2}}{1-z}\right)
(A_{\text{R}}+A_{\text{V}})  &  =\ln\left(  \frac{1}{\delta}\right)
(A_{\text{R}}+A_{\text{V}})+\text{subleading}\\
&  \approx\left(  -y_{\text{max}}+\frac{\Delta y}{2}+\ln\left(  \frac{Q}%
{k_{T}}\right)  \right)  (A_{\text{R}}+A_{\text{V}}).
\end{align}
Provided we stay within the soft-collinear region in which both the soft and
collinear approximations are valid, the $y_{\text{max}}$ dependence will
cancel with that coming from the soft contribution in Eq.(\ref{eq:softcol})
leaving only the logarithm. The leading effect of treating properly the
collinear region is therefore simply to introduce an effective upper limit to
the integration over rapidity in Eq.(\ref{eq:rap}). More precisely, we can
therefore estimate the leading behaviour simply by setting $y_{\text{max}%
}=\Delta y/2+\ln(Q/k_{T})$ in the soft integral, effectively including the
entire soft-collinear region. We are left with
\begin{equation}
~\frac{2\alpha_{s}}{\pi}\int_{Q_{0}}^{Q}\frac{dk_{T}}{k_{T}}\int
\limits_{Y/2}^{\ln(Q/k_{T})+\Delta y/2}\frac{dy~d\phi}{2\pi}~\omega_{1}=~\frac
{2\alpha_{s}}{\pi}\frac{1}{2}\ln^{2}(Q/Q_{0})+\text{subleading}.
\end{equation}
This is the super-leading logarithm: the failure of the `plus prescription'
has resulted in the generation of an extra collinear logarithm. The
implications for the gaps-between-jets cross-section are clear: collinear
logarithms can be summed into the parton density functions only up to scale
$Q_{0}$ and the logarithms in $Q/Q_{0}$ from further collinear evolution must
be handled separately. Moreover, since we now have a source of double
logarithms, the calculation of the single logarithmic series necessarily
requires knowledge of the two-loop evolution matrices \cite{Aybat:2006wq}.

Indeed we appear to have uncovered a breakdown of QCD coherence: radiation at
large angles does appear to be sensitive to radiation at low angles. However
this striking conclusion was arrived at under the assumption that it is
correct to order successive emissions in transverse momentum. Coherence
indicates that one does not need to take too much care over the ordering
variable, e.g. $k_{T}$, $E$ and $k_{T}^{2}/E$ are all equally good ordering
variables but the super-leading logarithms arise counter to the expectations
of coherence and in particular as a result of radiation which is both soft and
collinear. It is therefore required to prove the validity of $k_{T\text{ }}$
ordering before we can claim without doubt the emergence of super-leading
logarithms or confirm their size.

We note that the super-leading logarithm makes its appearance at the lowest
possible order in the perturbative expansion, i.e.\ at order $\alpha_{s}^{4}$
relative to the Born cross-section. More explicitly, the $O(\alpha_{s})$ and
$O(\alpha_{s}^{2})$ corrections to the Born cross-section simply never involve
more than one $i\pi$ term and hence any $i\pi$ terms must cancel since the
cross-section is real. The first candidate order at which two factors of
$i\pi$ can appear is therefore $O(\alpha_{s}^{3})$. However, the addition of
the gluon with the lowest $k_{T}$ can never generate a net factor of $i\pi$
since any such factors must cancel between the two diagrams where the lowest
$k_{T}$ gluon lies either side of the cut. Hence we anticipate that the first
super-leading logarithm makes a contribution%
\begin{equation}
\sigma\sim\sigma_{0}\left(  \frac{2\alpha_{s}}{\pi}\right)  ^{4}\ln^{5}\left(
\frac{Q}{Q_{0}}\right)  \pi^{2}Y. \label{eq:sllform}%
\end{equation}
Note that for each factor of $\pi$ we pay a price in colour (the leading
contribution in colour goes like $(\alpha_{s}N)^{n}$). The factor of $Y$ is
from the rapidity volume of the in-gap gluon. To be a little more explicit, we
now expand in $\alpha_{s}$.

The lowest order contribution for a single emission outside of the gap (with
$y>0$) is%
\begin{align}
\sigma_{1,\text{LO}}  &  =\sigma_{0~}\left(  \frac{2\alpha_{s}}{\pi}\right)
^{2}\frac{1}{8}\ln^{2}\left(  \frac{Q}{Q_{0}}\right)  \left\{  2Y~\int
\limits_{\text{out}}dy~~\frac{d\phi}{2\pi}~\omega_{24}\right. \nonumber\\
&  -\int\limits_{\text{out}}dy~~\frac{d\phi}{2\pi}\rho(Y,2\left\vert
y\right\vert )~\left[  \left(  N^{2}-2\right)  (\omega_{23}+\omega
_{14})-\omega_{13}+2(\omega_{12}+\omega_{34})-\omega_{24}\right] \nonumber\\
&  +\left.  \int\limits_{\text{out}}dy~\frac{d\phi}{2\pi}\lambda\left[
\left(  N^{2}-2\right)  \omega_{23}-\omega_{13}+2\omega_{34}\right]  \right\}
. \label{eq:leadingNGL}%
\end{align}
The dominant contributions at large enough $Y$ come from emissions close to
the edge of the gap. To see this we note that at large $Y$ the integral over
$\omega_{24}$ vanishes as $\exp(-2Y)$ and so the only significant contribution
arises from the terms proportional to $\rho$ and $\lambda$ which are dominated
by the region around $y=Y/2$.

The lowest order contribution that contains a super-leading logarithm is%
\begin{equation}
\sigma_{1,\text{SLL}}=-\sigma_{0}\left(  \frac{2\alpha_{s}}{\pi}\right)
^{4}\ln^{5}\left(  \frac{Q}{Q_{0}}\right)  \pi^{2}Y~\frac{(3N^{2}-4)}{480}.
\label{eq:leadingSLL}%
\end{equation}
Subsequent terms alternate in sign and are $\sim\alpha_{s}^{n}L^{n+1}\pi
^{2}N^{2}Y(NY)^{n-4}$ for large $N$ and $Y$.

Since we have only considered one emission out of the gap region, we should
convince ourselves that there is no possibility that the new collinear
logarithm cancels with a similar contribution from two (or more) emissions
outside of the gap. We here present an argument which confirms that the lowest
order (in $\alpha_{s}$) super-leading logarithm has nothing to cancel against.
As we have seen, this contribution occurs at order $\alpha_{s}^{4}$ relative
to the Born cross-section. We know that the gluon with the smallest $k_{T}$
does not give any $i\pi$ term and we know that there is an exact cancellation
if this gluon is outside of the gap. We also know that there is an exact
cancellation if all $i\pi$ terms are zero. Since the cross-section is real, we
must have an even number of $i\pi$ terms, which can in this lowest order case
only be two. Pulling all this together, we therefore have four gluons of which
the lowest $k_{T}$ gluon must lie inside the gap and two are Coulomb gluons.
Therefore we can only have, at most, one gluon outside of the gap.

Thus, we have shown that at order $\alpha_{s}^{4}$ all contributions are of
the type `zero gluons outside the gap' or `one gluon outside the gap' and we
have explicitly computed these and know that there is no cancellation.

As we have already shown, the miscancellation is specifically related to the
exchange of Coulomb gluons, since with the resulting $i\pi$ terms set to zero
cancellation is restored. It is worth recalling the special role of Coulomb
gluons in the proofs of factorization by Collins, Soper and Sterman
\cite{Collins88},\cite{Collins:1985ue},\cite{Collins:1998ps}. They consider
the exchange of potentially factorization-breaking soft gluons and show that
the eikonal gluons cancel in the sum over cuts through a given diagram, while
some Coulomb gluon terms remain uncancelled. Only after summing over all
diagrams in which a Coulomb gluon is exchanged, in particular including
diagrams in which it is attached to the hadron remnants, can the corresponding
contribution be shown to cancel. In our case, since we consider a high-$p_{t}$
process (the exchanged gluons we are interested in populate the
strongly-ordered region $k_{T}\gg Q_{0}$ where we assume $Q_{0}\gg
\Lambda_{\text{QCD}}$), emission from the hadron remnants is irrelevant
(power-suppressed) and hence we have no \textit{a priori} reason to assume
that the Coulomb phase terms will cancel. It must be checked explicitly and in
our case they do not.

\section{Numerical results\label{sec:results}}

In the following figures\footnote{We generically write the in-gap
cross-section as $\sigma_{0}$ and the out-of-gap cross-section $\sigma_{1}$.},
we have computed the out-of-gap cross-section obtained by summing
Eq.(\ref{eq:real}) and Eq.(\ref{eq:virtual}) each evaluated in the
super-leading (soft and collinear) approximation. This amounts to setting all
the $\omega_{ij}=0$ except $\omega_{12}=\omega_{13}=\omega_{14}=1$ (in the
case $y>0\,$) and $\rho(Y,2\left\vert y\right\vert )=\lambda=0$. In addition,
the integral over rapidity is performed over an interval of size $\ln
(Q/k_{T})$ and we multiply by a factor of 2 to account for the possibility
that the out-of-gap gluon can be either side of the gap. We refer to the
cross-section thus computed as `SLL' in all of the plots since it contains the
complete super-leading contribution. For comparison, we also compute the sum
of Eq.(\ref{eq:real}) and Eq.(\ref{eq:virtual}) without making the collinear
approximation. In this case the integral over $y$ is over the region
$Y/2<|y|<\Delta y/2+\ln(Q/k_{T})$ and we take $R=1$. These cross-sections are
labelled `all' in the plots and they necessarily include a partial summation
of the single logarithmic terms as well as the super-leading terms. Throughout
we keep the strong coupling fixed at $\alpha_{S}=0.15$ and our cross-sections
are usually normalized to the fully resummed cross-section corresponding to
zero gluons outside of the gap region, i.e.\ as determined by Eq.(\ref{eq:OS}%
). \begin{figure}[h]
\begin{center}
\centerline{\includegraphics[width=10cm]{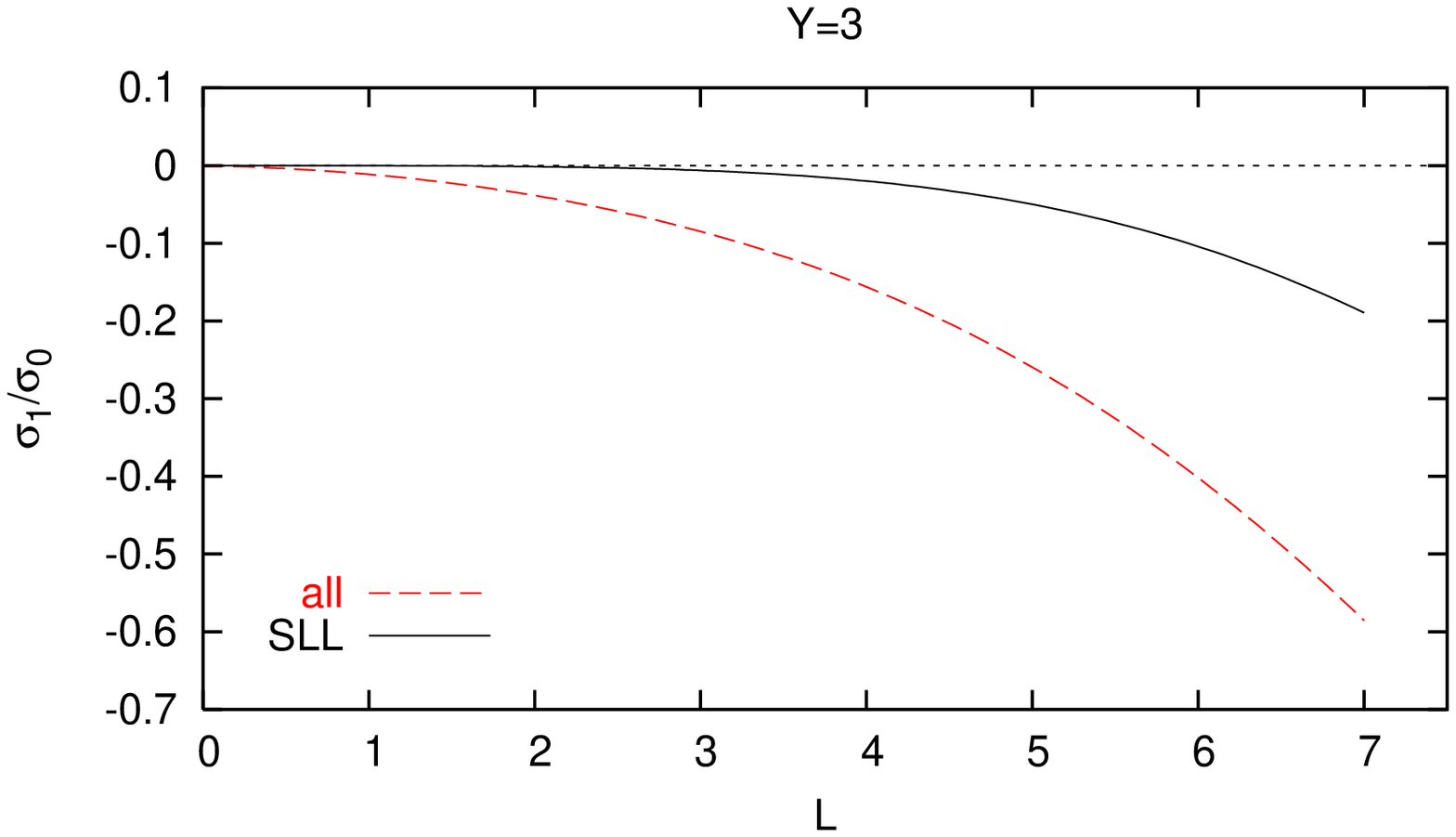}}\centerline{\includegraphics[width=10cm]{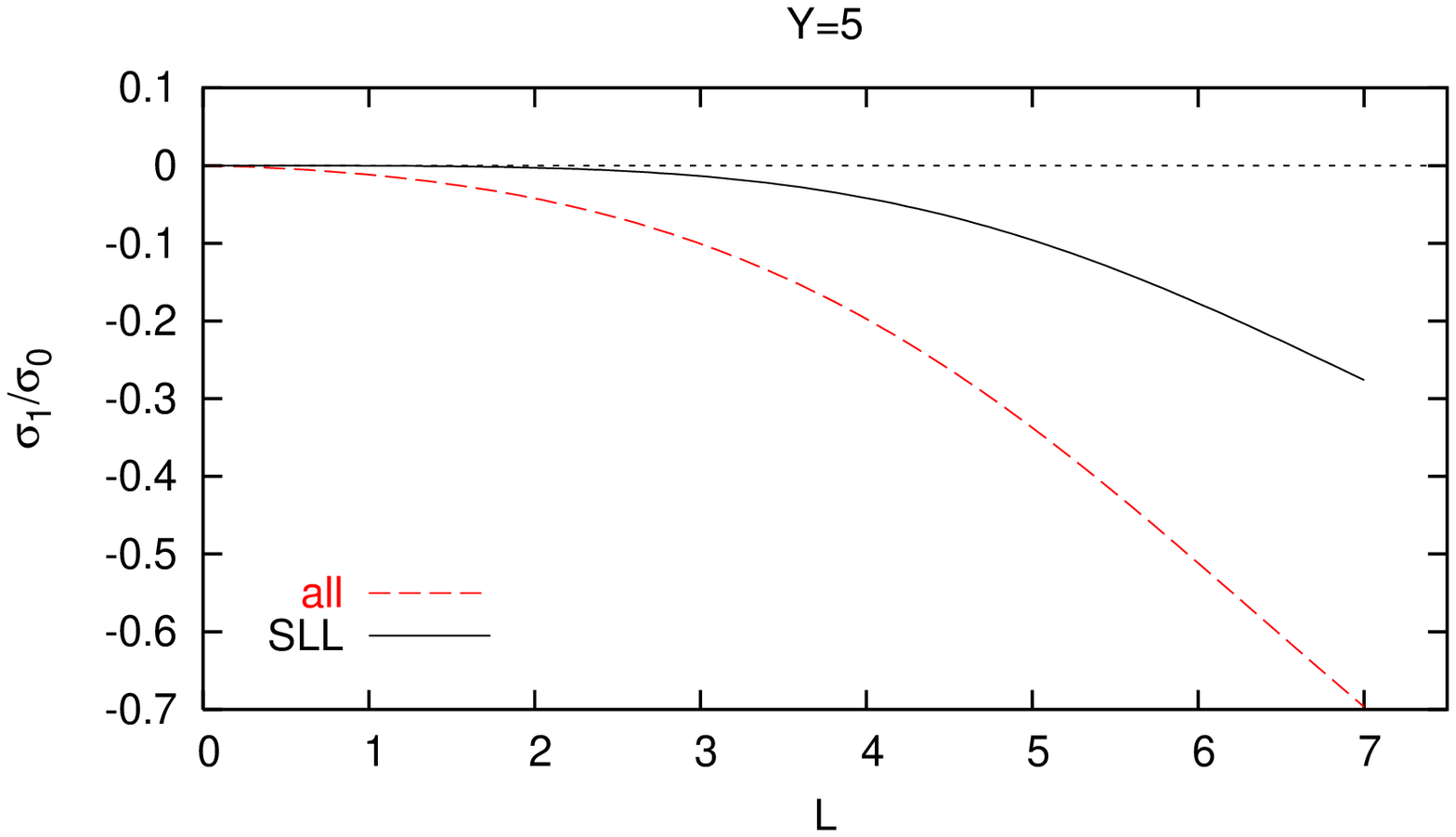}}
\end{center}
\caption{$L$ dependence of the out-of-gap cross-section (normalized to the
in-gap cross-section) at two different values of $Y$.}%
\label{fig:Ldep}%
\end{figure}\begin{figure}[h]
\begin{center}
\centerline{\includegraphics[width=10cm]{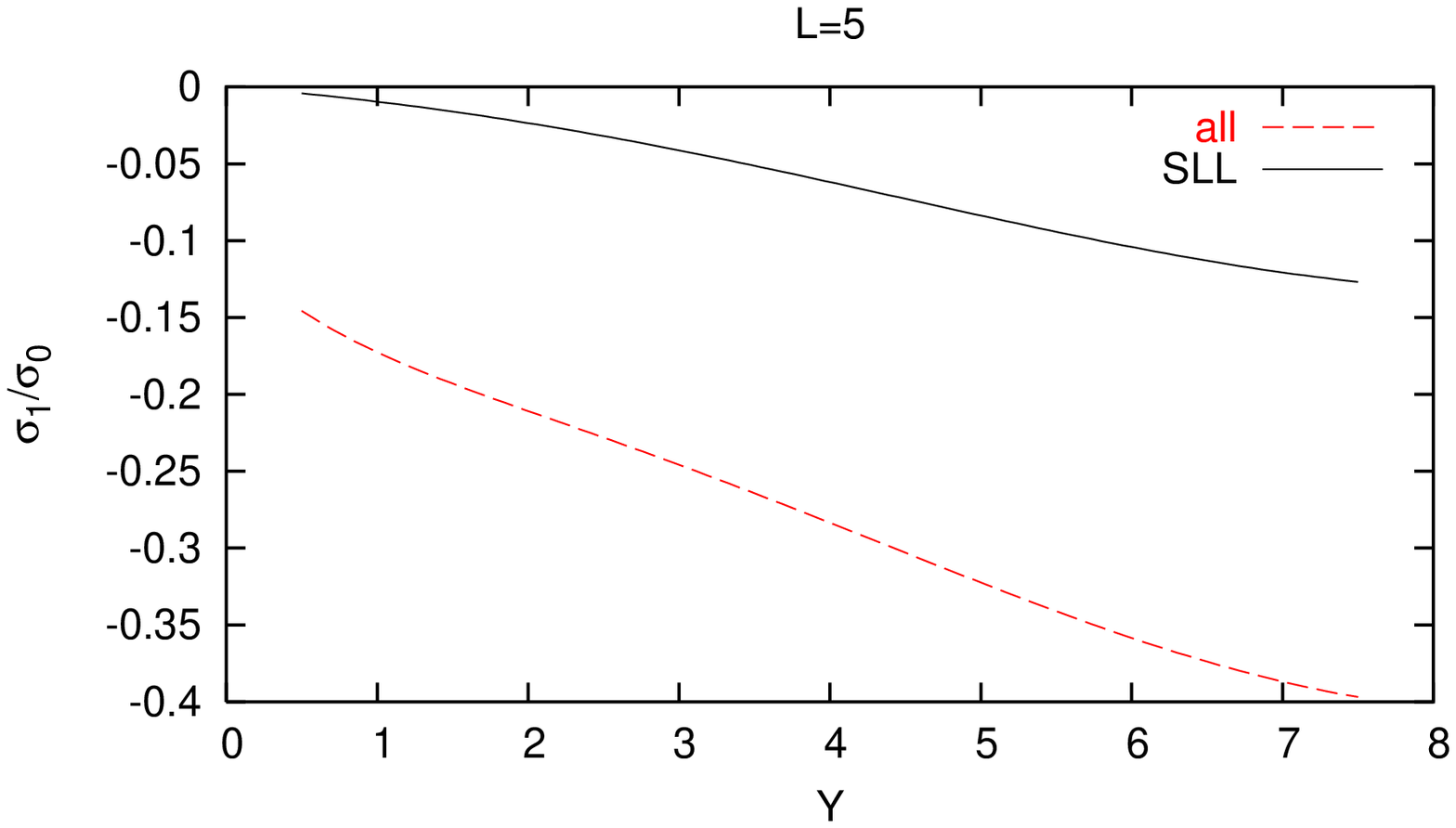}}\centerline{\includegraphics[width=10cm]{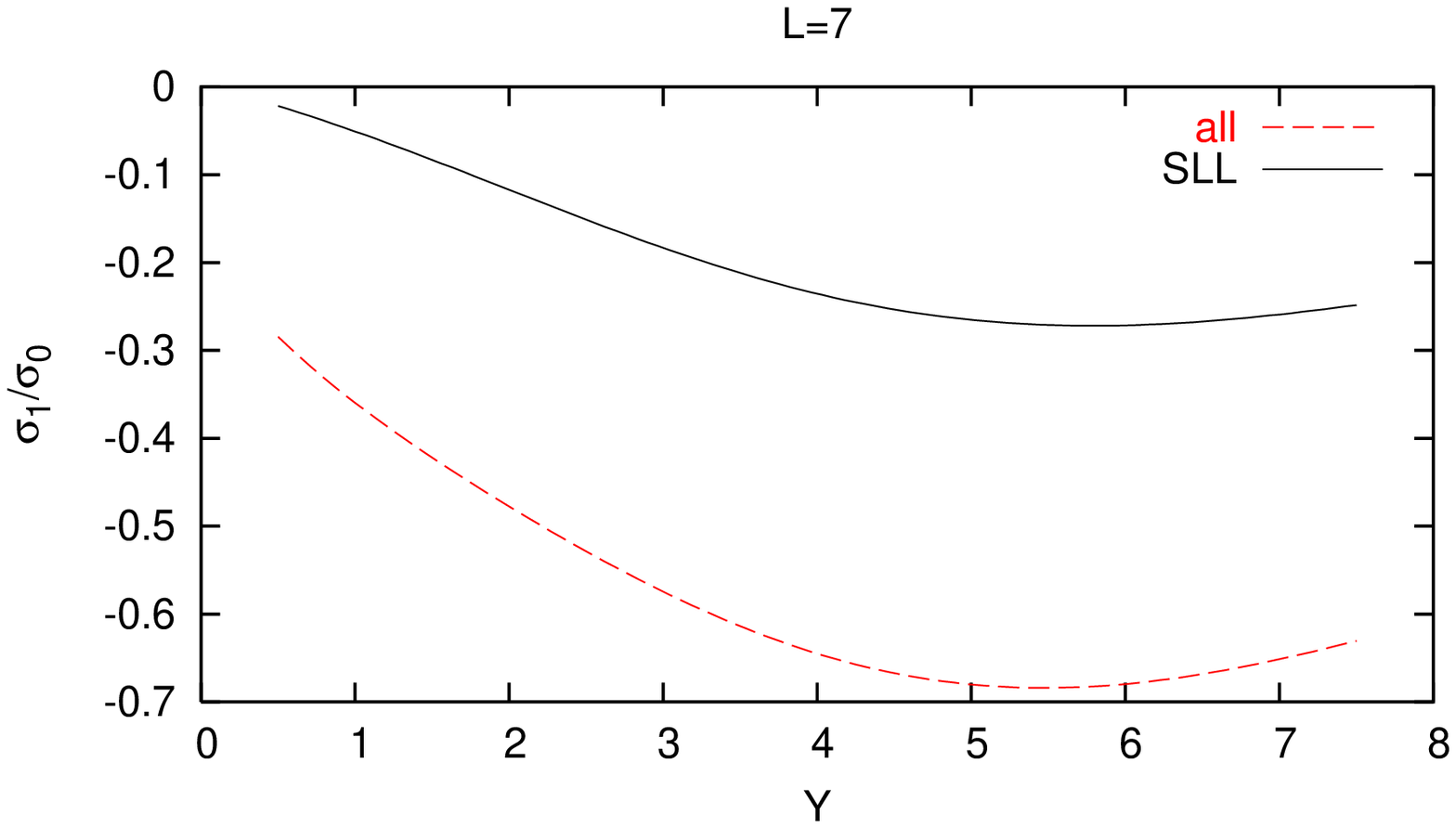}}
\end{center}
\caption{$Y$ dependence of the out-of-gap cross-section (normalized to the
in-gap cross-section) at two different values of $L$.}%
\label{fig:Ydep}%
\end{figure}

The plots in Fig.\ref{fig:Ldep} show the cross-section dependence upon
$L=\ln(Q^{2}/Q_{0}^{2})$ at two different values of $Y$ whilst the dependence
upon $Y$ at two different values of $L$ is illustrated in Fig.\ref{fig:Ydep}.
\ It seems that while the out-of-gap cross-section is not dominant anywhere it
is also not negligible. This is of course already known:\ the non-global
logarithms are generally significant. We can also see from these plots that
the super-leading series is generally small relative to the `all' result for
$L\lesssim4$, which indicates that the single logarithms are
phenomenologically much more important than the formally super-leading logs at
these values of $L$. Of course one should remember that our calculations are
for the emission of one gluon outside the gap region and the full
super-leading series requires the computation of any number of such gluons.

\begin{figure}[h]
\begin{center}
\centerline{\includegraphics[width=10cm]{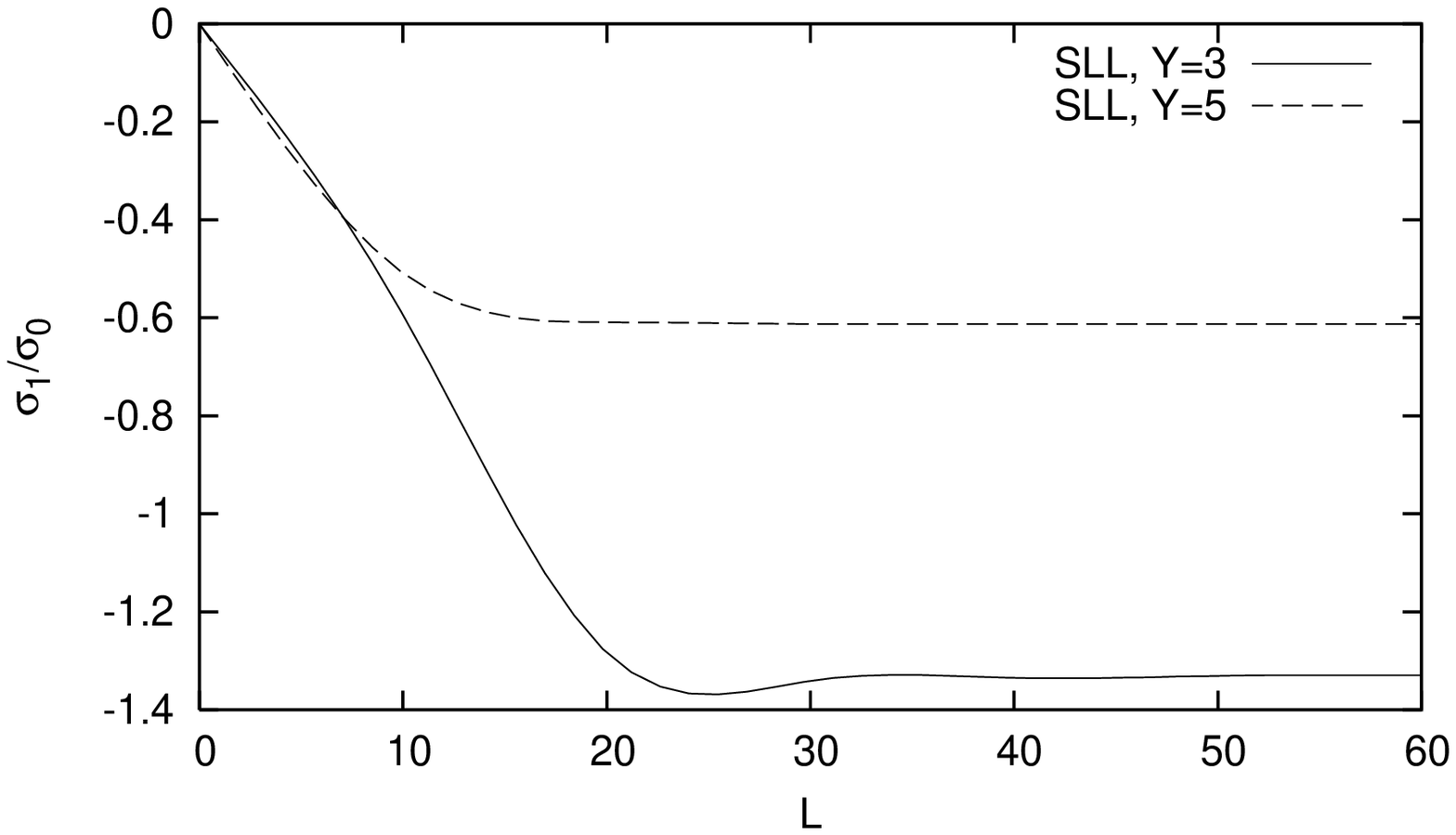}}
\end{center}
\caption{$L$ dependence of the out-of-gap cross-section (normalized to the
in-gap cross-section) at two different values of $Y$ and plotted out to very
large $L$.}%
\label{fig:largeLdep}%
\end{figure}

\begin{figure}[h]
\begin{center}
\includegraphics[width=10cm]{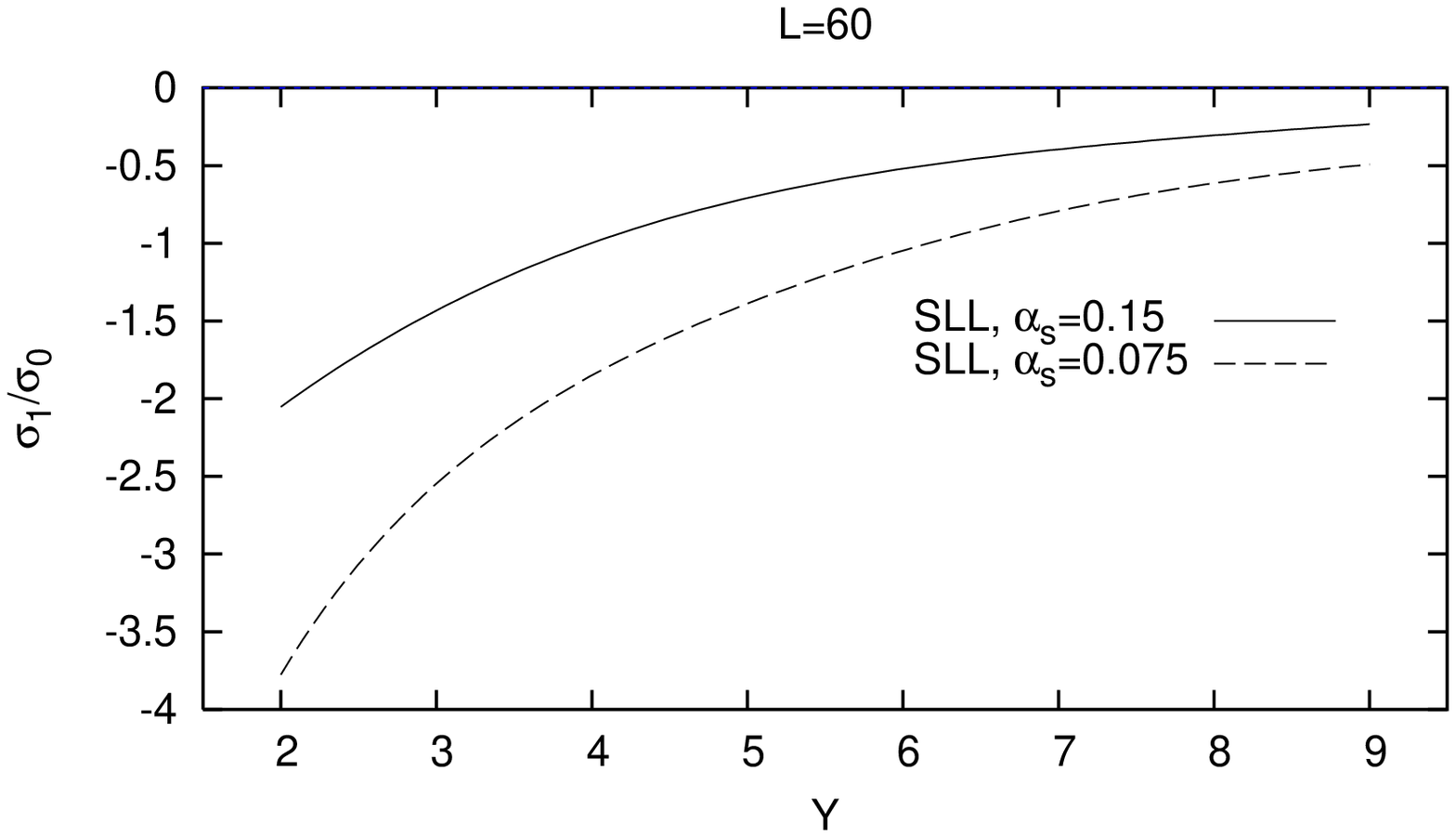}\\[0pt]
\end{center}
\caption{The $Y$ dependence of the large $L$ behaviour of the out-of-gap
cross-section normalized to the in-gap cross-section at two different values
of $\alpha_{s}$.}%
\label{fig:satL}%
\end{figure}

\begin{figure}[h]
\begin{center}
\centerline{\includegraphics[width=10cm]{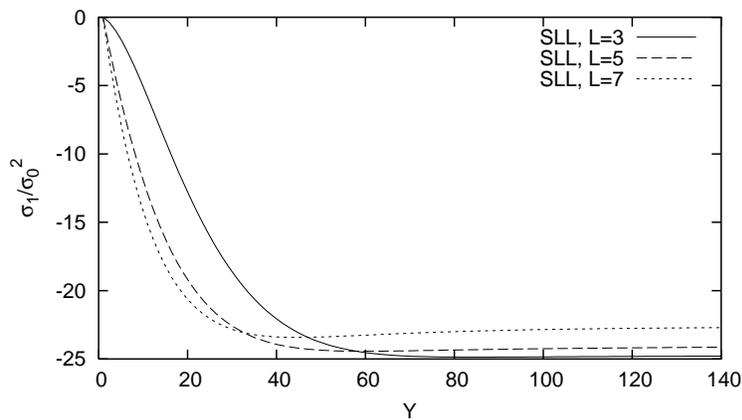}}
\end{center}
\caption{$Y$ dependence of the out-of-gap cross-section normalized to the
square of the in-gap cross-section at three different values of $L$ and
plotted out to very large $Y$.}%
\label{fig:largeYdep}%
\end{figure}

\begin{figure}[ptb]
\begin{center}
\includegraphics[width=10cm]{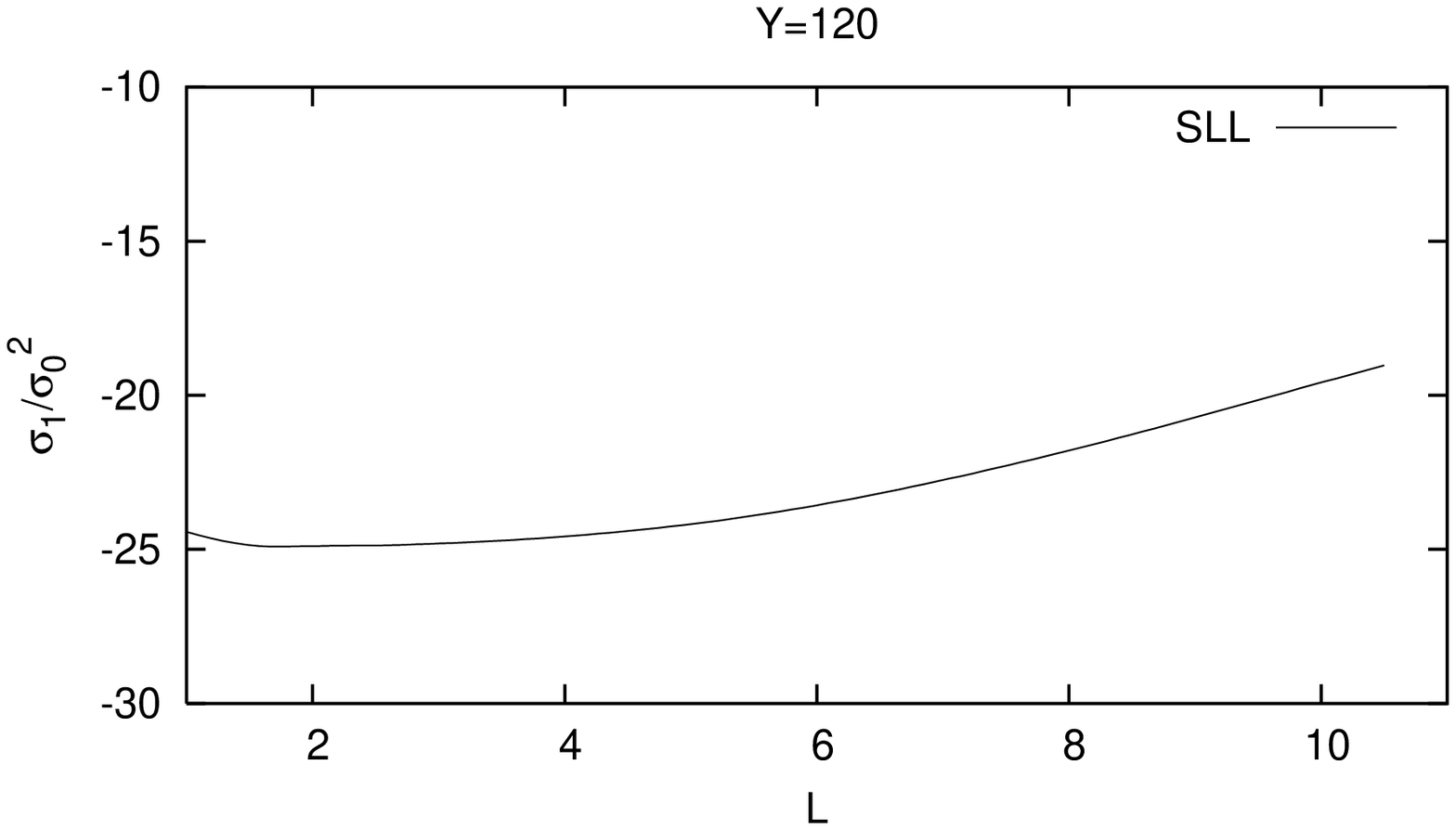}\newline
\end{center}
\caption{The $L$ dependence of the large $Y$ behaviour of the out-of-gap
cross-section normalized to the square of the in-gap cross-section}%
\label{fig:satY}%
\end{figure}

From a more theoretical perspective it is interesting to take a look at the
cross-sections out to larger values of $L$ and $Y$. In Fig.\ref{fig:largeLdep}
we show the cross-section out to large values of $L$. It is immediately
striking that the cross-section asymptotes to a constant value, which implies
that the out-of-gap cross-section is directly proportional to the in-gap
cross-section at large $L$ with a $Y$ dependent prefactor. In
Fig.\ref{fig:satL} we show how the large $L$ behaviour of the cross-section
varies with $Y$.

Fig.\ref{fig:largeYdep} shows the dependence of the cross-section out to large
$Y$. Note that this time we have normalized the cross-section by the square of
the in-gap cross-section (and have set the Born cross-section equal to unity).
The cross-section again saturates at large enough $Y$. Fig.\ref{fig:satY}
shows how the large $Y$ behaviour varies with $L$. We consider the fact that
$\sigma_{1}\sim-\sigma_{0}^{2}$ at large $Y$ to reflect the deeper link which
is known to exist between QCD dynamics in non-global observables and small-$x$
physics where such non-linear effects lie behind the phenomenon of parton
saturation \cite{Marchesini:2003nh}--\cite{Weigert:2005us}.

\section{Conclusions}

Conventional calculations of non-global observables assume that emission well
away from the region in which the observable is calculated cancels. When
starting this work, we aimed to check this assumption for one of the simplest
non-global observables in hadron--hadron collisions, the gaps-between-jets
cross-section, by explicitly calculating the all-orders contribution from
configurations with one gluon outside the gap region. Based on the pioneering
work of Dasgupta and Salam, we expected to find additional contributions from
emission \textit{just }outside the gap. Physically, the probability that such
radiation is not accompanied by additional nearby radiation reduces the gap
cross-section, giving rise to additional towers of leading logarithms; the
so-called non-global logs. We indeed found such a contribution, illustrated in
Eq.(\ref{eq:leadingNGL}).

However, when calculating the evolution of five-parton configurations produced
by real radiation outside the gap, we found a mismatch between it and the
evolution of the four-parton configurations corresponding to virtual emission.
This can be traced to the Coulomb phase terms (the imaginary parts of the loop
integrals) coming from singularities that pinch the contour integral at the
point $k_{+}=k_{-}=0$. As illustrated in Eq.~(\ref{eq:ipito0}), if these terms
are artificially set to zero, the mismatch vanishes. However, keeping these
terms, a mismatch remains, even for emission arbitrarily far away from the gap
region. Integrating over phase space results in a new \emph{superleading}
logarithm, formally more important than any so-called leading logarithm
previously included, as illustrated in Eq.(\ref{eq:leadingSLL}). Our
conclusions are subject to the caveat that we have assumed the validity of
transverse momentum ordering for successive soft gluon emissions.

Although from our numerical results it may appear that the phenomenological
impact of this formally-dominant effect is modest for $L$ and $Y$ values of
interest, we should recall that we have only calculated the contribution from
one gluon outside the gap. Having identified such a contribution, it is
clearly necessary to examine how contributions from arbitrary numbers of
gluons outside the gap will contribute. In fact we see no reason why the
argument at the end of Section \ref{sec:SLL} should not hold for the leading
such contribution and expect that at the $n$th order of perturbation theory
the leading contribution will come from $n-3$ gluons outside the gap,
resulting in a term $\sim\alpha_{s}^{n}L^{2n-3}\pi^{2}Y$. Calculating such
contributions analytically seems a formidable challenge without a deeper
understanding of the colour evolution of multi-parton systems.

We close this paper with a remark about the more theoretical interest of our
result. One can view the gaps-between-jets process as a look at the pomeron
loop in QCD, since one sums over radiation outside the gap (corresponding to a
cut pomeron) and forbids it inside the gap (corresponding to one pomeron
either side of the cut). In \cite{Forshaw:2005sx} we calculated the
conventional gap-between-jets cross-section in the high energy limit and
showed that it is equivalent to the BFKL result in the region in which both
are valid. In this paper, we noted that in the high-energy (large $Y$) limit
the cross section for one emission outside the gap is proportional to the
square of the conventional gap cross-section, offering a tantalizing clue to
the structure of higher orders. A deeper understanding of this connection
would almost certainly open new avenues to understanding non-global observables.

\section*{Acknowledgements}

We thank Mrinal Dasgupta and Gavin Salam for many interesting discussions.
This work was supported by a grant from the UK's Particle Physics and
Astronomy Research Council.

\appendix{}

\section{The block diagonal basis\label{app2}}

Here we present the action of shifting from the $t$-channel colour basis to
the block diagonal basis. We first exploit the fact that we can add any
imaginary multiple of the unit matrix to the evolution matrices without
affecting any observables in order to introduce
\begin{equation}
\mathbf{\Lambda}^{\prime}=\mathbf{\Lambda}+\frac{N}{4}i\pi\mathbf{1.}%
\end{equation}
The required block diagonalization of $\mathbf{\Lambda}^{\prime}$ is effected by%

\begin{equation}
\mathbf{R}=\sqrt{\frac{N}{2(N^{2}-1)}}\left(
\begin{array}
[c]{cccc}%
\frac{1}{2}s_{y} & -\frac{1}{2}s_{y} & s_{y} & \frac{1}{2N}s_{y}\\
\frac{N}{N+2}s_{y} & \frac{N}{N-2}s_{y} & 0 & s_{y}\\
-\frac{1}{2} & \frac{1}{2} & 1 & -\frac{1}{2N}\\
1 & 1 & 0 & -1
\end{array}
\right)  .
\end{equation}
The real emission matrix then transforms to%

\begin{align}
&  \mathbf{D}^{\mu}\rightarrow\mathbf{R}^{-1}\mathbf{D}^{\mu}=\nonumber\\
&  \frac{1+s_{y}}{2}\sqrt{\frac{N^{2}-1}{2N}}\left(
\begin{array}
[c]{cc}%
\frac{N+2}{N+1}(h_{4}^{\mu}-h_{2}^{\mu}) & \frac{1}{2}\frac{N+2}{N+1}%
(h_{4}^{\mu}-h_{1}^{\mu}+\frac{1}{N}h_{2}^{\mu}-\frac{1}{N}h_{4}^{\mu})\\
\frac{N-2}{N-1}(h_{2}^{\mu}-h_{4}^{\mu}) & \frac{1}{2}\frac{N-2}{N-1}%
(h_{4}^{\mu}-h_{1}^{\mu}+\frac{1}{N}h_{4}^{\mu}-\frac{1}{N}h_{2}^{\mu})\\
h_{3}^{\mu}-h_{1}^{\mu} & \frac{1}{2N}(h_{4}^{\mu}-h_{2}^{\mu})\\
\frac{2N}{N^{2}-1}(h_{4}^{\mu}-h_{2}^{\mu}) & \frac{1}{N^{2}-1}\left(
h_{1}^{\mu}-h_{3}^{\mu}+N^{2}\left(  h_{3}^{\mu}-h_{2}^{\mu}\right)
+2(h_{2}^{\mu}-h_{4}^{\mu})\right)
\end{array}
\right) \nonumber\\
&  -\frac{1-s_{y}}{2}\sqrt{\frac{N^{2}-1}{2N}}\left(
\begin{array}
[c]{cc}%
\frac{N+2}{N+1}(h_{3}^{\mu}-h_{1}^{\mu}) & \frac{1}{2}\frac{N+2}{N+1}%
(h_{3}^{\mu}-h_{2}^{\mu}+\frac{1}{N}h_{1}^{\mu}-\frac{1}{N}h_{3}^{\mu})\\
\frac{N-2}{N-1}(h_{1}^{\mu}-h_{3}^{\mu}) & \frac{1}{2}\frac{N-2}{N-1}%
(h_{3}^{\mu}-h_{2}^{\mu}+\frac{1}{N}h_{3}^{\mu}-\frac{1}{N}h_{1}^{\mu})\\
h_{4}^{\mu}-h_{2}^{\mu} & \frac{1}{2N}(h_{3}^{\mu}-h_{1}^{\mu})\\
\frac{2N}{N^{2}-1}(h_{3}^{\mu}-h_{1}^{\mu}) & \frac{1}{N^{2}-1}\left(
h_{2}^{\mu}-h_{4}^{\mu}+N^{2}\left(  h_{4}^{\mu}-h_{1}^{\mu}\right)
+2(h_{1}^{\mu}-h_{3}^{\mu})\right)
\end{array}
\right)
\end{align}
and the evolution matrix becomes%
\begin{align}
\mathbf{\Lambda}\mathbf{\rightarrow R}  &  ^{-1}\mathbf{\Lambda}^{\prime
}\mathbf{R=}\left(
\begin{array}
[c]{cccc}%
{\normalsize \lambda}_{1} & 0 & {\normalsize 0} & {\normalsize 0}\\
{\normalsize 0} & \lambda_{2} & {\normalsize 0} & {\normalsize 0}\\
{\normalsize 0} & {\normalsize 0} & \frac{N^{2}-1}{4N}\rho(Y,\Delta y) &
\frac{N^{2}-1}{4N^{2}}i\pi\\
{\normalsize 0} & {\normalsize 0} & {\normalsize i\pi} & -\frac{1}{N}%
i\pi+\frac{N}{2}Y+\frac{N^{2}-1}{4N}\rho(Y,\Delta y)
\end{array}
\right) \nonumber\\
&  +\left(
\begin{array}
[c]{cccc}%
N & 0 & 0 & 0\\
0 & N & 0 & 0\\
0 & 0 & N & 0\\
0 & 0 & 0 & N
\end{array}
\right)  \frac{1}{4}\rho(Y,2\left\vert y\right\vert )\nonumber\\
&  +\left(
\begin{array}
[c]{cccc}%
1 & 0 & 0 & 0\\
0 & -1 & 0 & 0\\
0 & 0 & -N & 0\\
0 & 0 & 0 & -N
\end{array}
\right)  \frac{\lambda}{4}%
\end{align}
where the $\lambda_{i}$ are specified in Eq.(\ref{eq:eigenvals}).

Finally, the colour matrix transforms to
\begin{equation}
\mathbf{S}_{R}\mathbf{\rightarrow R}^{\dag}\mathbf{S}_{R}\mathbf{~R}=\left(
\begin{array}
[c]{cccc}%
\frac{N^{2}}{2}\frac{N+1}{N+2} & 0 & 0 & 0\\
0 & \frac{N^{2}}{2}\frac{N-1}{N-2} & 0 & 0\\
0 & 0 & N^{2} & 0\\
0 & 0 & 0 & \frac{1}{4}(N^{2}-1)
\end{array}
\right)  .
\end{equation}

\end{document}